\documentclass[aps,superscriptaddress,nofootinbib,a4paper,reprint]{revtex4-1}
\usepackage{graphicx}
\usepackage{amssymb}
\usepackage{amsthm}
\usepackage{amsfonts}
\usepackage{amsmath}
\usepackage{subcaption} 
\usepackage{xspace}
\newcommand{\Finesse}{\textsc{Finesse}\xspace}
\newcommand{\Comsol}{\textsc{Comsol}\xspace}
\usepackage[dvipsnames]{xcolor}

\begin{document}
\title[PIs at LIGO]{The Influence of Dual-Recycling on Parametric Instabilities at Advanced LIGO}

\author{A. C. Green}
\email{agreen@star.sr.bham.ac.uk}
\author{D. D. Brown}
\author{M.\ Dovale-\'Alvarez}
\author{C. Collins}
\author{H. Miao}
\author{C. M. Mow-Lowry}
\author{A. Freise}
\affiliation{University of Birmingham, UK}
\date{\today}
\begin{abstract}
Laser interferometers with high circulating power and suspended optics, such as the LIGO gravitational wave detectors, experience an optomechanical coupling effect known as a \textit{parametric instability}: the runaway excitation of a mechanical resonance in a mirror driven by the optical field. This can saturate the interferometer sensing and control systems and limit the observation time of the detector. Current mitigation techniques at the LIGO sites are successfully suppressing all observed parametric instabilities, and focus on the behaviour of the instabilities in the Fabry-Perot arm cavities of the interferometer, where the instabilities are first generated. In this paper we model the full dual-recycled Advanced LIGO design with inherent imperfections. We find that the addition of the power- and signal-recycling cavities shapes the interferometer response to mechanical modes, resulting in up to four times as many peaks. Changes to the accumulated phase or Gouy phase in the signal-recycling cavity have a significant impact on the parametric gain, and therefore which modes require suppression. 
\end{abstract}

\maketitle

\section{Parametric Instability in Complex Interferometers}
Prior to the first detection of a gravitational wave signal in September 2015 \cite{gw150914}, the two LIGO detectors went through the first stage of a major upgrade, which aimed to boost their sensitivity by a factor of 10 \cite{Harry10}. One major aspect of this upgrade was a significant increase in optical power circulating in the arm cavities, which is expected to reduce shot noise and therefore improve detector sensitivity. 
During the first observational run (O1) of Advanced LIGO, circulating power approaching 100~kW was consistently achieved; the target power is 750~kW resulting in a design sensitivity of 200~Mpc binary neutron star range \cite{gw150914Dets}.

In this paper we discuss a phenomenon known as \textit{parametric instabilities} \cite{BSV01}, a consequence of using high optical power in the interferometer.  Parametric instabilities result from an interaction between the radiation pressure of the optical field and the natural vibrational modes of the mirrors. In the presence of positive optical feedback this can couple energy from the field into the mirror mode, resulting in exponential growth of the mechanical oscillation.

Parametric instabilities (PIs) have now been observed in prototype optical cavities \cite{Corbitt2006, Zhao15, Chen15}, and at LIGO \cite{Evans15}, where the mechanical modes have been observed to ring up until the interferometer control systems failed. Current mitigation strategies are focused on technologies already built into the interferometers: using ring heaters to change the optical gain of problematic higher order optical modes \cite{Zhao05, Gingin06, Degallaix07, Zhao08, Susmithan12}, and using electrostatic drivers to actively damp the mechanical mode that is unstable \cite{Miller2011788, CBlair16}. So far these techniques are proving successful; however, as the circulating power is increased towards the design level the severity and number of PIs will increase, resulting in more unstable modes.

Over many years we have developed a detailed simulation model using \Finesse \cite{finesse20} to describe behaviours in the Advanced LIGO detectors \cite{Hild07, Sorazu13, Somiya16}. In this paper, we present numerical analyses of PIs in the full design configuration of Advanced LIGO \cite{AdvancedLIGO15}. This complements existing extensive analytical and numerical modelling of PIs \cite{Gurkovsky07, Evans2010665, Gras2010-b, Kells06, Danilishin14, Zhao15}.

First, we present an overview of how \Finesse is used to model PIs throughout this work. In section~\ref{sec:Builddowns} we study the parametric gain of specific mechanical modes and how this changes when recycling cavities are introduced, including inherent defects such as astigmatism. Section~\ref{sec:SRMphi} then focuses on parameters of the signal-recycling cavity and consequences for the parametric gain in a realistic interferometer configuration. 
We find that changes to the tuning or accumulated Gouy phase of the signal-recycling cavity have a significant impact on parametric gain, and therefore which modes will require suppression. However whether this has consequences for the current mitigation scheme is not yet known. 

\subsection{Numerically Modelling PIs}

\Finesse \cite{finesse20} is a fast, frequency-domain interferometer simulation tool. 
It is particularly suited to modelling parametric instabilities as it easily provides the required mechanical-to-optical transfer functions in imperfect and arbitrary interferometer configurations using Hermite-Gaussian beams. 
Previously this has been used to apply limits to the number and type of higher order modes used in simulation \cite{Bond11}, and investigate the potential use of higher order Laguerre-Gauss modes to reduce thermal noise in future gravitational wave detector designs \cite{Carbone13}. It is also actively used in LIGO commissioning and design modelling \cite{Somiya16, Brooks15}. \Finesse and its Python wrapper PyKat \cite{Pykat} are open source and freely available for others to use in future studies.

Parametric instabilities can be considered as a feedback system \cite{Evans2010665} resulting from the linear interaction between the optical field and a vibrational mode of a suspended mirror within the interferometer. The figure of merit for determining the stability of a vibrational mode in an interferometer is called the \textit{parametric gain}, $\mathbb{R}$, where $\mathbb{R} \geq 1$ corresponds to an instability. Typically this is evaluated by calculating individual optical transfer functions of higher order optical modes through the interferometer and then summing the effect of each of these optical modes, as described in Appendix~\ref{app:PIintro}. 

The open loop transfer function for a motion back onto itself is $T(\Omega) = \frac{p(\Omega)}{\Delta p(\Omega)}$, where $p(\Omega)$ describes the amplitude motion spectrum of a mechanical mode at a frequency $\Omega$. This exists in \Finesse as a diagonal element in the inverted interferometer matrix. A single sparse matrix solution can then be used to evaluate $T(\Omega)$ at the frequency of the $m$th mechanical mode, $\omega_m$. Using the frequency-domain equation of state, we find that the parametric gain of the $m$th mechanical mode, $\mathbb{R}_m$, can then be directly extracted from the real part of this transfer function, as described in \cite{phd.brown2015}:

\begin{equation}
\mathbb{R}_m= 1- \Re\Big({\frac{1}{T(\omega_m)}}\Big)
\end{equation}

Since \Finesse computes the light field amplitudes of all Hermite-Gauss (HG) modes to a specified order, $T(\omega_m)$ contains components from all these optical fields directly. In order to calculate the spatial overlap between the optical fields and the mechanical modes, \Finesse must be supplied with a \textit{surface motion map}. Typically these maps are produced using a finite element modelling package, which computes the mechanical resonant modes of the bulk optic. From the bulk modes, the normalised motion of the front surface and the corresponding resonant frequency can be extracted and used as inputs into \Finesse models. Further details about using finite element modelling tools with \Finesse are provided in chapter 3 of \cite{phd.brown2015}.

\subsection{General Method}\label{sec:Method}

\begin{figure}[htb]
\begin{center}
\includegraphics[width=0.45\textwidth]{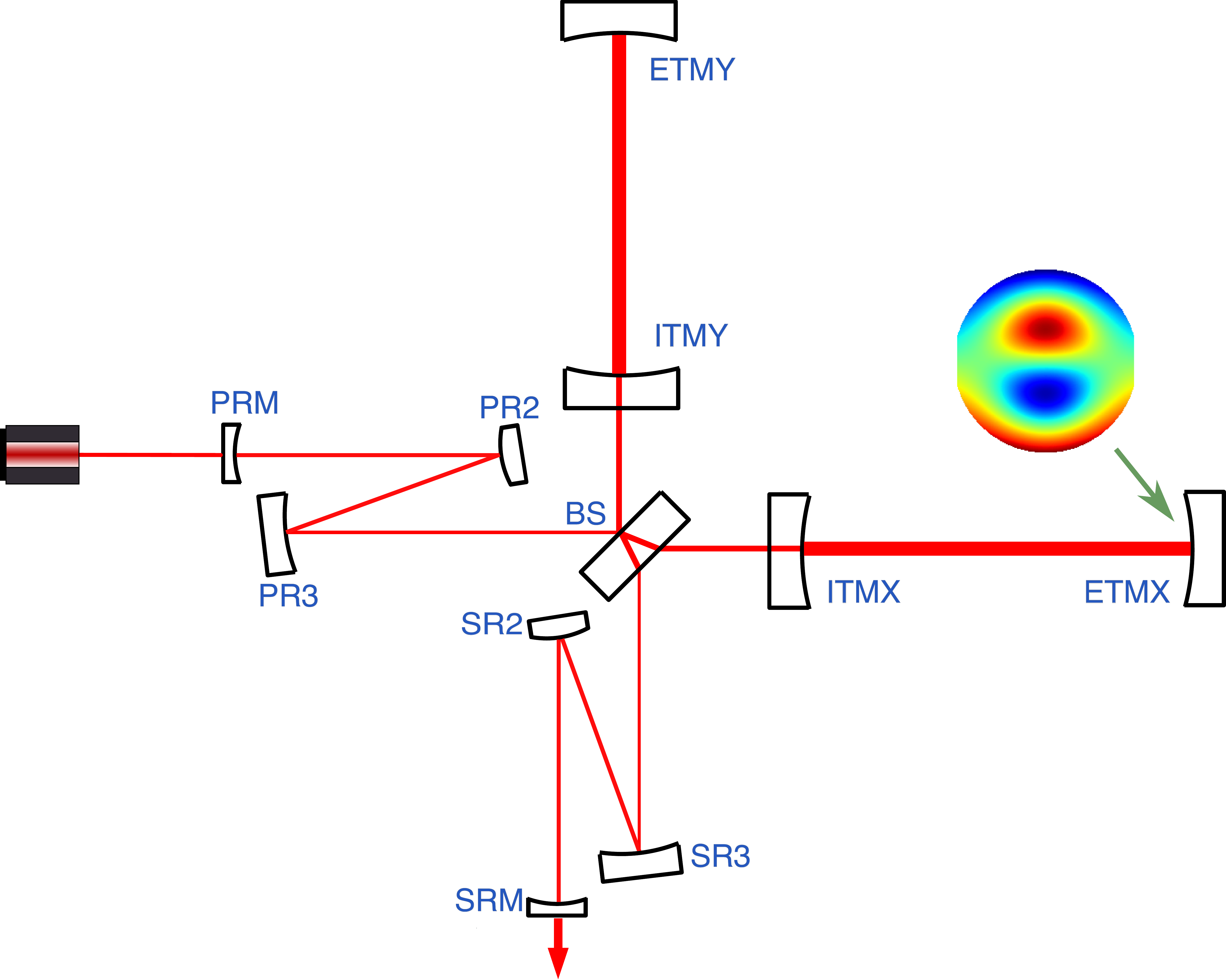}
\caption{Optical Layout used throughout these simulations, indicating the location of the applied surface motion map. The X- and Y-arm cavities are formed by the Input and End Test Masses (I- and ETMs), and together form a Fabry-Perot Michelson via the beamsplitter (BS). The Power Recycling Cavity (PRC) is formed between the arm cavities and the Power Recycling Mirror (PRM), via Power Recycling cavity mirrors PR2 and PR3. The Signal Recycling Cavity (SRC) is similarly formed between the arm cavities and Signal Recycling Mirror (SRM) via Signal Recycling cavity mirrors SR2 and SR3.}
\label{fig:OptLayout}
\end{center}
\end{figure}

\begin{table}[h!]
\centering
\begin{tabular}{c|r|r}
 Cavity 	&  Property 			& Value \\ \hline\hline
 X-arm 	& $f_{\rm msf}$ 		& 5.155~kHz \\
		& $f_{\rm pole}$		& 42.34~Hz \\ \hline
 Y-arm 	& $f_{\rm msf}$			& 5.155~kHz \\
		& $f_{\rm pole}$		& 42.34~Hz \\ \hline
 PRX 	& x-axis $f_{\rm msf}$ 	& 377.0~kHz \\
 		& y-axis $f_{\rm msf}$ 	& 358.3~kHz \\
		& $f_{\rm pole}$ 		& 309.5~kHz \\ \hline
PRY		& x-axis $f_{\rm msf}$ 	& 377.5~kHz \\
 		& y-axis $f_{\rm msf}$ 	& 359.0~kHz \\
		& $f_{\rm pole}$ 		& 310.0~kHz \\ \hline
 SRX 	& x-axis $f_{\rm msf}$ 	& 288.5~kHz \\
 		& y-axis $f_{\rm msf}$ 	& 255.3~kHz \\
		& $f_{\rm pole}$ 		& 420.5~kHz \\ \hline
 SRY 	& x-axis $f_{\rm msf}$ 	& 288.9~kHz \\
 		& y-axis $f_{\rm msf}$ 	& 255.6~kHz \\
		& $f_{\rm pole}$ 		& 421.0~kHz \\ \hline
\end{tabular}
\caption{Key frequencies derived from the Advanced LIGO design model. $f_{\rm pole}$ is the pole frequency, describing the linewidth (half-width-half-maximum) of the cavity.  $f_{\rm msf}$ is optical higher order mode separation frequency. In cavities where the beam is astigmatic, the mode separation frequency differs between the x- and y-axes. PRX(Y) and SRX(Y) refer to the cavities formed between the PRM or SRM and ITMX(Y) respectively (see figure~\ref{fig:OptLayout}). }\label{tab:CavProperties}
\end{table}

A complete model of the core optics in Advanced LIGO forms the basis of the simulation, as depicted in figure~\ref{fig:OptLayout}. This model uses design parameters given in \cite{AdvancedLIGO15}. Key frequencies derived from this design are listed in Table~\ref{tab:CavProperties}. 

The detector is based on a Michelson interferometer. Fabry-Perot cavities in the arms, formed by the Input- and End Test Masses (I- and ETMs) are used to amplify gravitational wave signals, and the Power Recycling Cavity formed between the Power Recycling Mirror (PRM) and arm cavities increases the circulating power to decrease shot noise. The Signal Recycling Cavity, formed by the Signal Recycling Mirror (SRM) and arm cavities, can be tuned to amplify or resonantly extract signal sidebands; currently the LIGO detectors operate in this Dual Recycled configuration using Resonant Sideband Extraction (RSE). 

As in the design, the X- and Y-arm cavities are identical. Parameters within this core model, such as mirror positions, angles and curvatures, may then be varied to change the response of the interferometer, and consequently the parametric gain. We have used feedback loops, mimicking those used at the detector sites, to check that these parameter sweeps do not move the model away from an \textit{operating point} for the interferometer linear degrees of freedom.  This means that both arm cavities and the power-recycling cavity are resonant for the carrier field, and the inner Michelson (formed by the beamsplitter and Input Test Masses) is tuned to a dark fringe on transmission. We refer to in-phase signals that are common to both arms and therefore reflected back towards the laser as `common' signals, while `differential' signals, with a 180$^\circ$ phase difference between arms, are transmitted from the beamsplitter to the detection port. 

\begin{figure}[htb]
\centering
    \begin{subfigure}[b]{0.27\textwidth}
        \includegraphics[width=\textwidth]{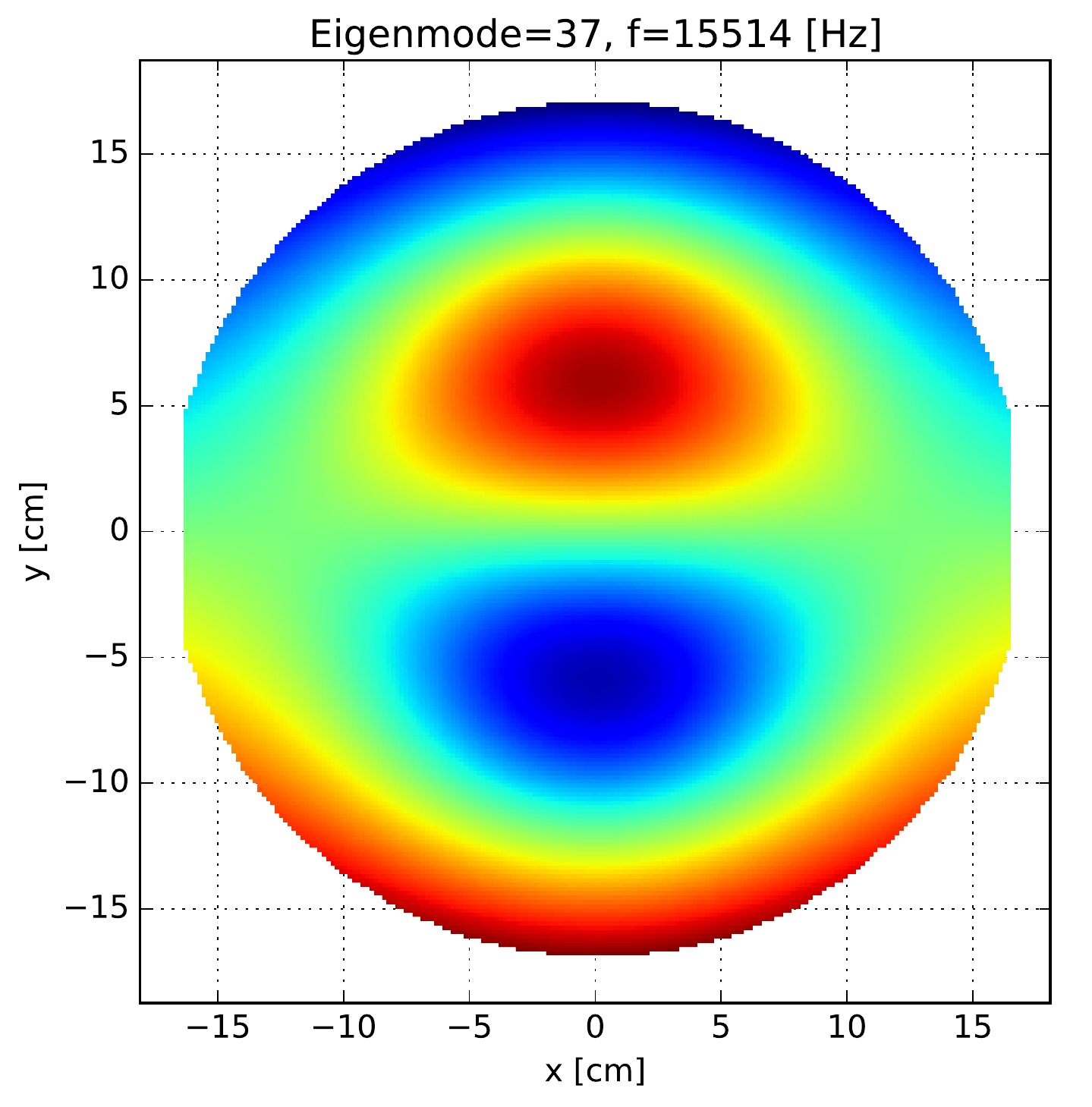}
        \caption{Mode 37}
        \label{fig:m37}
    \end{subfigure}
    ~ 
    \begin{subfigure}[b]{0.27\textwidth}
        \includegraphics[width=\textwidth]{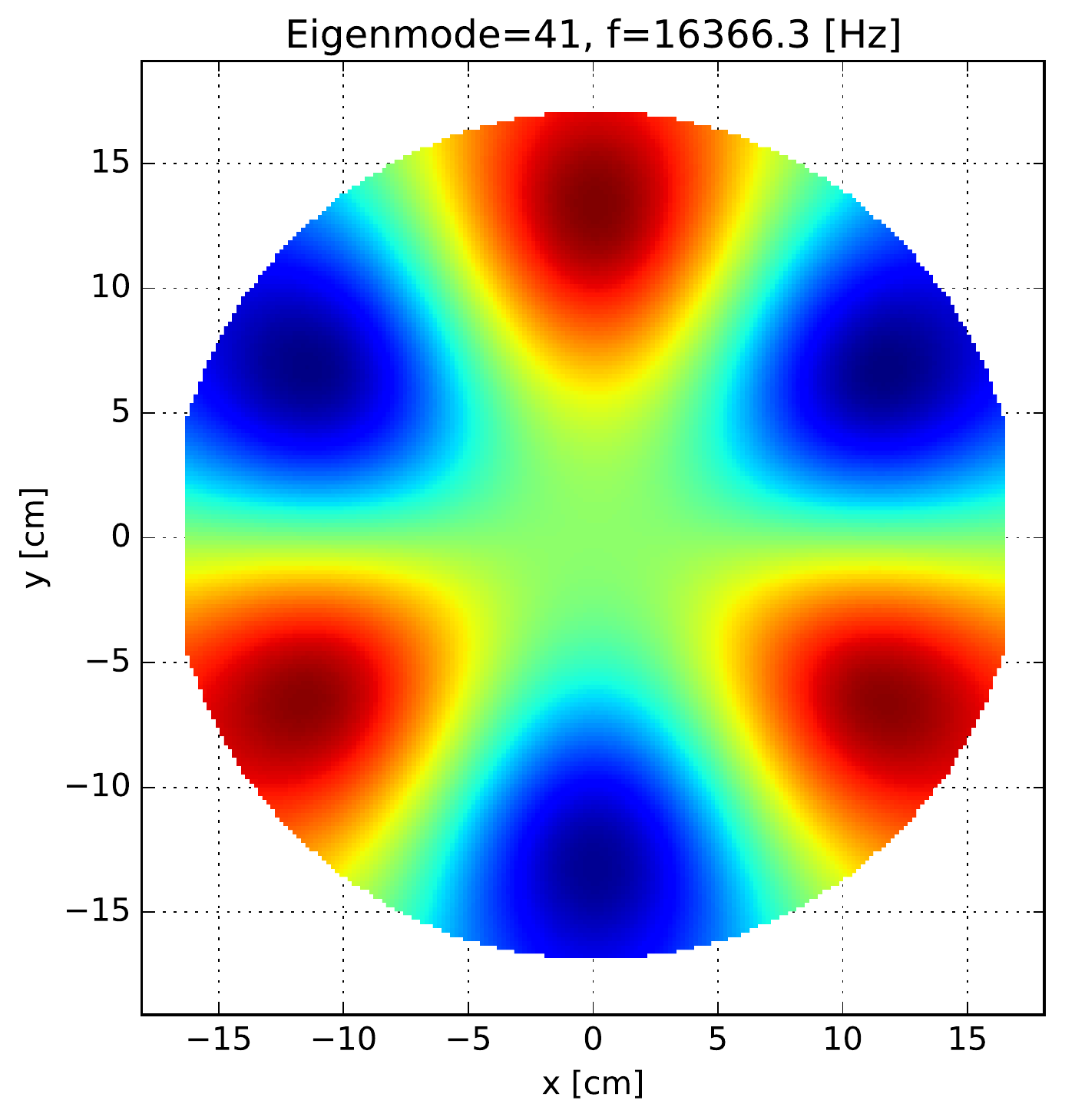}
        \caption{Mode 41}
        \label{fig:m41}
    \end{subfigure}
    ~ 
    \begin{subfigure}[b]{0.27\textwidth}
        \includegraphics[width=\textwidth]{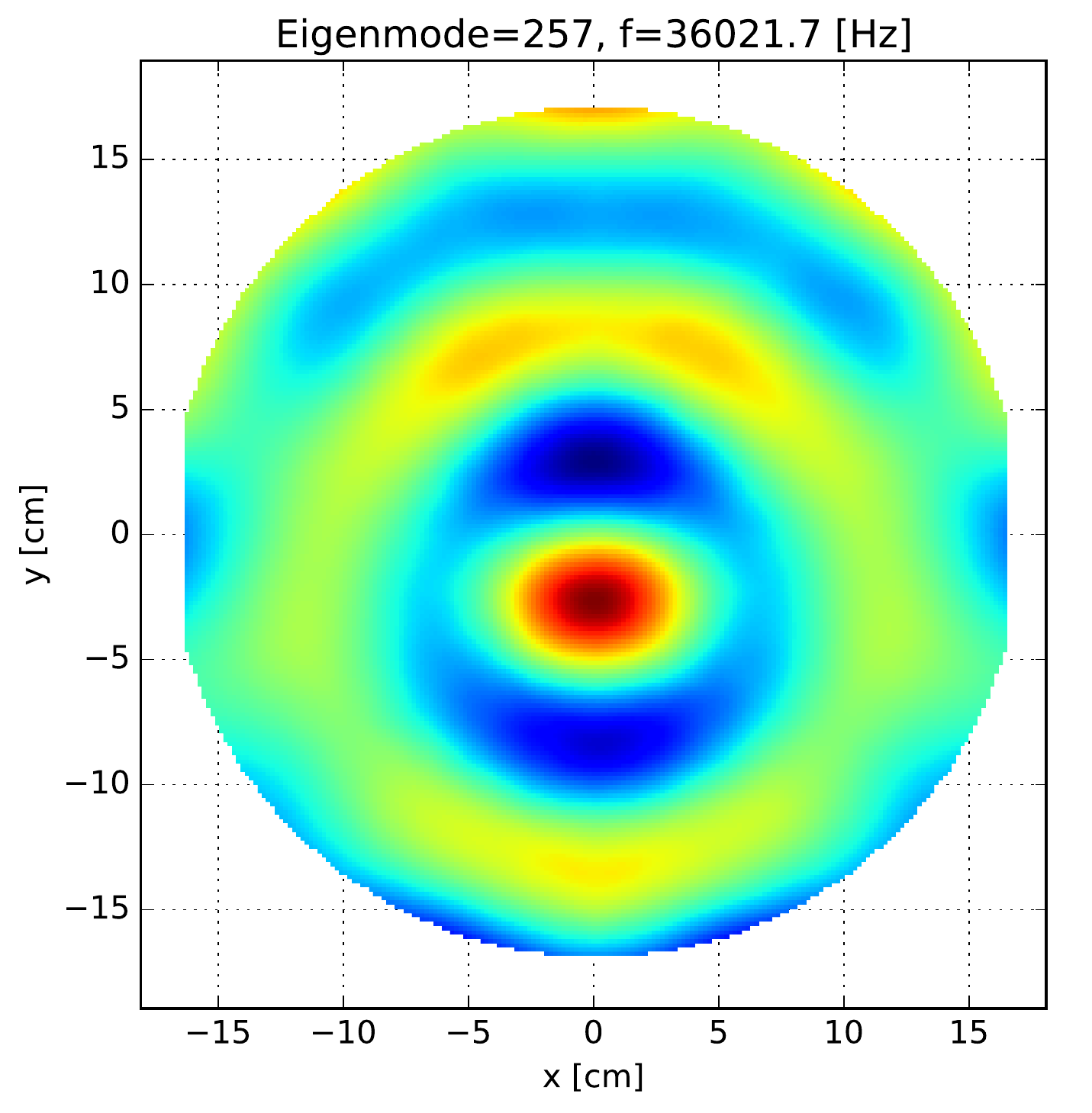}
        \caption{Mode 257}
        \label{fig:m257}
    \end{subfigure}
\caption{Surface motion maps for mechanical modes used throughout this modelling.
Mode 37 is associated with the first observed parametric instability at Advanced LIGO. Modes 41 and 257 are examples of mechanical modes that interact with the dual recycled interferometer differently to mode 37, as described in section ~\ref{sec:SRMphi}.}
\label{fig:maps}
\end{figure}

Surface motion maps for the Advanced LIGO mirrors were produced using \Comsol. Examples are shown in figure~\ref{fig:maps}, where listed mode numbers are those generated by \Comsol. In particular, mode 37 has strong spatial overlaps with HG03 and HG21 optical modes and is associated with the first observation of a PI in a LIGO detector \cite{Evans15}. This observation was made at the Livingston detector (which is dual recycled and tuned for resonant sideband extraction), operated with 50~kW circulating arm power resulting in a parametric gain of $\mathbb{R}$=2. Minor (0.15\%) adjustments have therefore been made to the radius of curvature of the four test masses in our model to reflect the observed resonant frequency and parametric gain of this mode. 

Each simulation of a parametric instability applies one surface motion map to the End Test Mass of the X-arm (ETMX), as shown in figure~\ref{fig:OptLayout}. The simulation also takes the resonant mechanical mode frequency and Q-factor as inputs; by default we use a Q-factor of $10^7$ and the resonant frequency computed by \Comsol. This means that we can explore the combined parameter space of mechanical mode frequency and interferometer parameters. Note that since we only study effects due to ETMX, modes from different test masses and any cross-coupling between these are not considered in this study. 

\section{Parametric instability in increasingly complex interferometers}\label{sec:Builddowns}

\begin{figure}[htb]
\begin{center}
\includegraphics[width=0.4\textwidth]{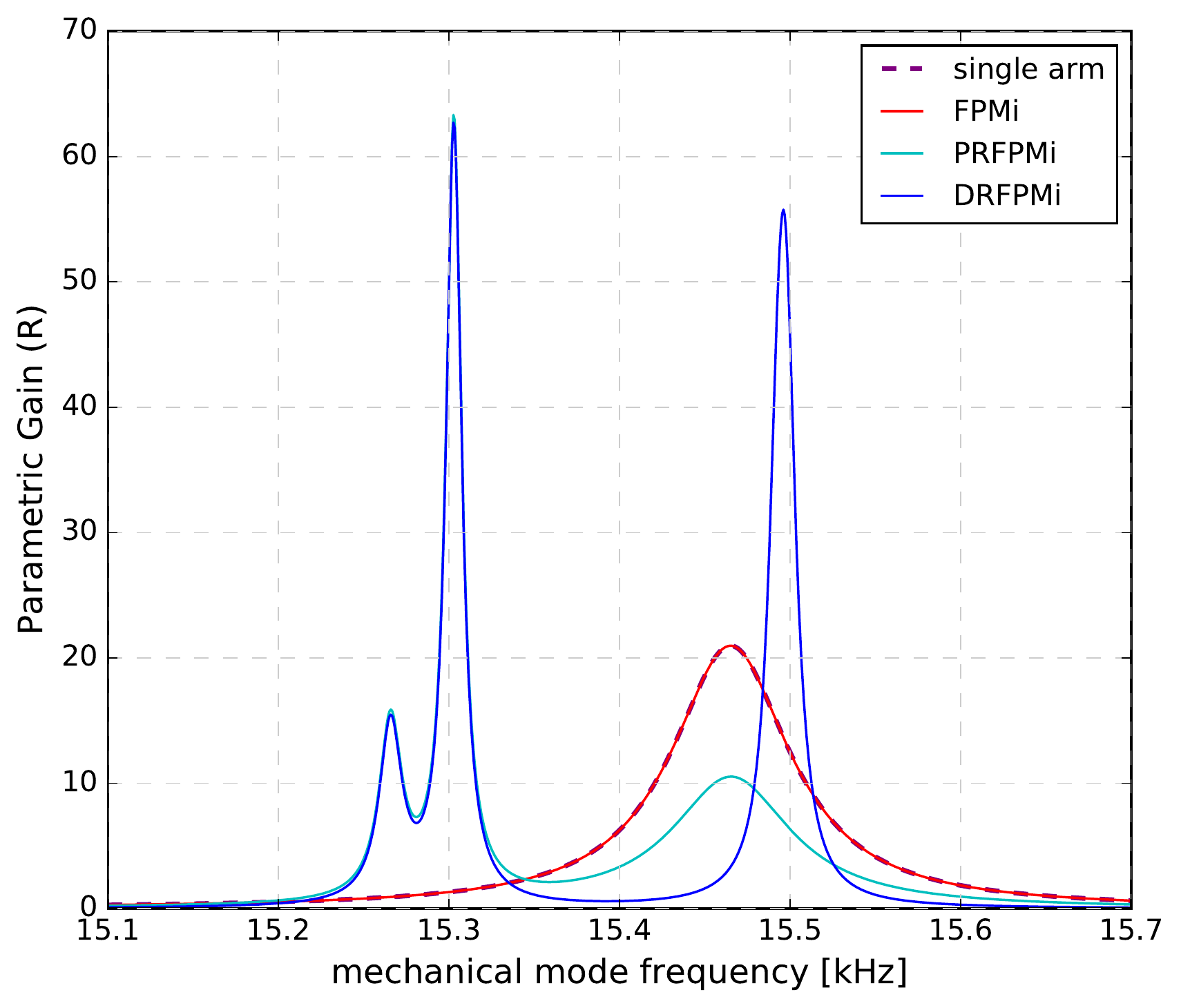}
\caption{Parametric gain of mode 37 as a function of mechanical mode frequency in the case of a single arm cavity, Michelson interferometer with Fabry-Perot arm cavities, Power-Recycled-, and Dual-Recycled Fabry-Perot Michelson. Arm circulating power is constant between all traces.}
\label{fig:builddown-as}
\end{center}
\end{figure}

Figure~\ref{fig:builddown-as} depicts the parametric gain of mode 37 (see figure~\ref{fig:maps}) as a function of the resonant frequency of the mechanical mode, using the method described in section~\ref{sec:Method}. We compare a single X-arm cavity to Michelson Interferometers with just Fabry-Perot arms (FPMi),  Power-Recycling (PRFPMi),  and Dual-Recycling (DRFPMi). We find that the presence of the power- and signal-recycling cavities significantly shapes the optical response and resulting parametric gain, in agreement with \cite{Evans2010665}. 
To allow direct comparison, the input power was adjusted to maintain a constant power circulating in the arm(s) in all cases. 

In both the single arm cavity and FPMi cases we see a typical single broad peak. This corresponds to an overlap between the mechanical mode frequency and the 5.16kHz mode separation frequency of the arm cavity, which allows a 3rd order optical mode to resonate.

Introducing the power-recycling mirror results in a cavity coupling between the X-arm and both the power-recycling cavity (PRC) and Y-arm. The condition for resonance is therefore complicated. We see the introduction of \textit{two} new peaks, since the PRC includes spherically curved mirrors at non-normal incidence, producing an astigmatic beam. This results in the HG03 and HG21 modes picking up different amounts of Gouy phase in the cavity. We describe these peaks as \textit{common} peaks due to their association with the reflection port of the Michelson. The frequency separation between these common peaks and the original single cavity peak is 182~Hz, while the separation between the two common peaks is 36~Hz. 

Similarly, adding the signal-recycling mirror produces an additional set of couplings via the signal-recycling cavity (SRC). This time the new resonance condition results in two \textit{differential} peaks, offset from the single cavity resonance by 30~Hz. These two peaks are unresolved due to the low finesse of the SRC, appearing as a broadening of the peak when compared to the non-astigmatic case. We also see that the original broad peak is supressed.

Our model allows us to treat the resonant frequency of each mechanical mode as a tuneable parameter, as discussed in section~\ref{sec:Method}. Changes of frequency on the scales explored here are not something we expect to see in reality, however plots of this kind are useful diagnostic tools. They allow us to explore the response of the interferometer to a mechanical mode, whose resonant frequency may shift and is unknown prior to measurement, without changing the interferometer state. 

Experimental work at both the Hanford and Livingston detectors has attributed a mirror motion with the shape of mode 37 to observed parametric instabilities at 15.53kHz. We see that in our model this falls within a differential peak in parametric gain, indicating that properties of the signal-recycling cavity could also be used to influence the gain of this mode in the interferometer, as shown in section~\ref{sec:SRMphi}. However, other modes will match different resonant conditions in the interferometer, for example resonating via the PRC. Improving the behaviour for one mechanical mode may worsen the situation for another. 

\begin{figure}[t]
\begin{center}
\includegraphics[width=0.4\textwidth]{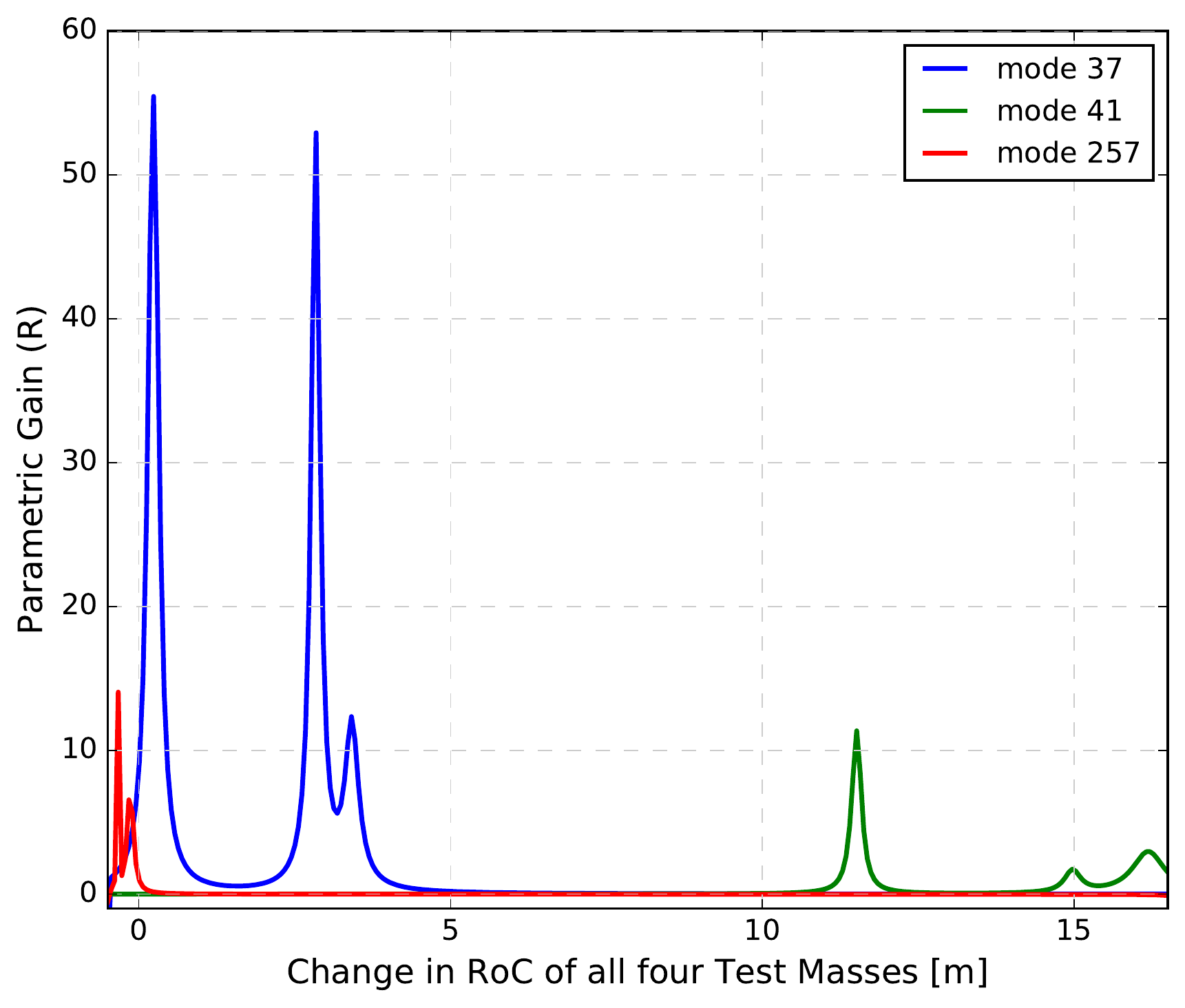}
\caption{The parametric gain of modes 37, 41 and 257 depend on the radii of curvature (RoC) of the test masses. In this plot, the RoC of all four test masses are changed by the same amount simultaneously. For context, the reference curvature of ETMX is 2248m.}
\label{fig:RvRoCs}
\end{center}
\end{figure}

The internal properties of the arm cavities can be used to suppress parametric instabilities. Unlike the SRC properties discussed below, changes to the radii of curvature (RoCs) of the test masses are known to influence parametric gain by altering the optical response. They are therefore one focus of efforts to suppress PIs at the detector sites. Figure~\ref{fig:RvRoCs} plots the parametric gain of modes 37, 41 and 257 on ETMX when the RoCs of all four test masses are increased simultaneously by the same amount. Once again we find four peaks in the trace for mode 37. We can also see that a significant change in RoC is expected to stabilise mode 37, but overcompensation could result in instability through mode 41. 

\section{The Signal-Recycling Cavity}\label{sec:SRMphi}

\subsection{Tuning}
In addition to Advanced LIGO's current broadband operation using resonant sideband extraction (RSE), the tuning of the signal-recycling cavity can be adjusted to produce an operational mode that is optimised for a particular gravitational wave source \cite{meers89}. In particular, an SRC detuning of $\phi =16^\circ$ is proposed for optimal binary neutron star detection \cite{AdvancedLIGO15}, where a tuning of $ 360^\circ$ corresponds to a mirror displacement of one wavelength. We find that the parametric gain of some mechanical modes has a strong dependence on the tuning of the signal-recycling cavity length.

\begin{figure}[t]
\begin{center}
\includegraphics[width=0.45\textwidth]{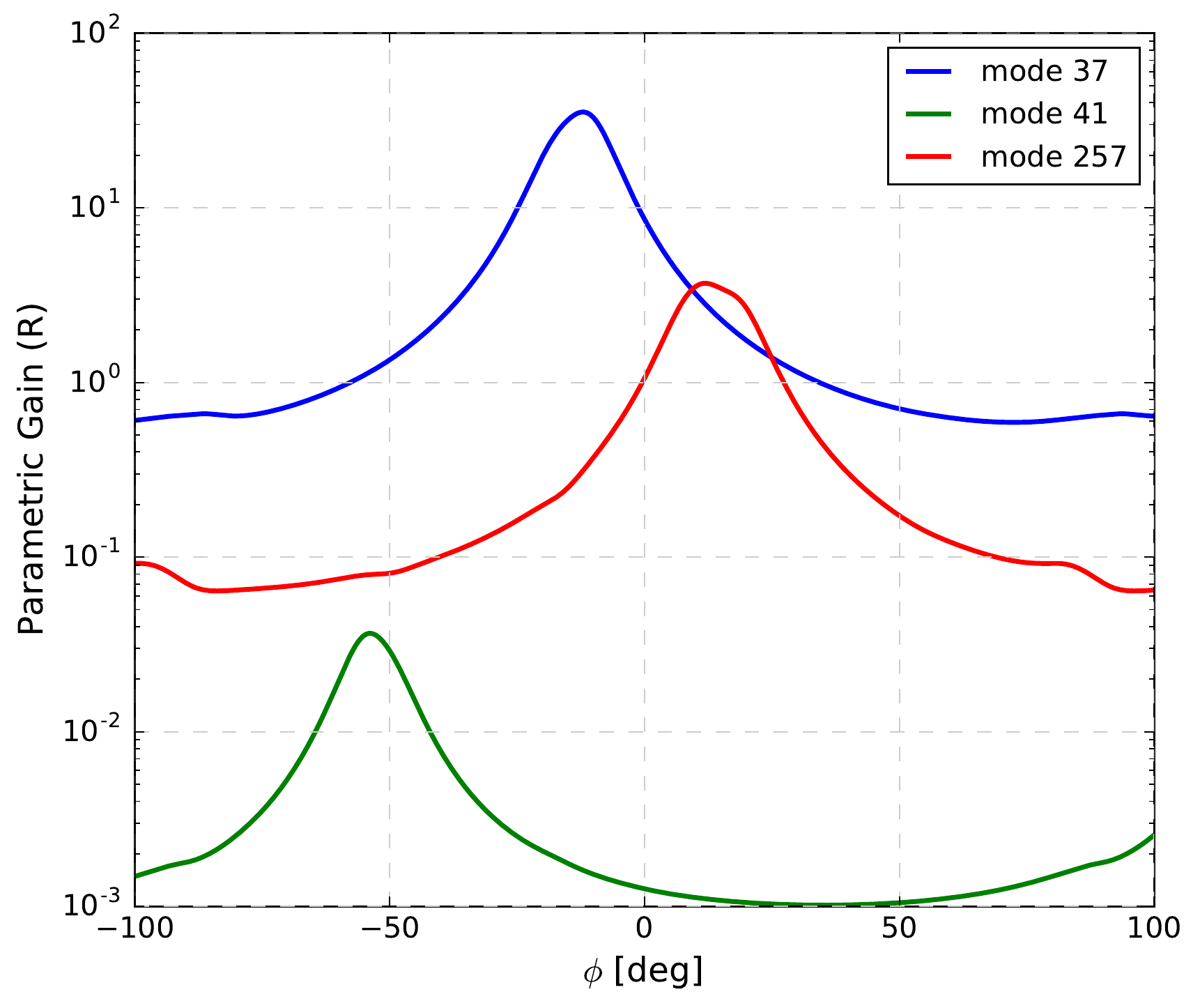}
\caption{The parametric gain of modes 37, 41 and 257 (see figure~\ref{fig:maps}) depend on the tuning of the signal-recycling cavity relative to RSE.} 
\label{fig:SRMphi}
\end{center}
\end{figure}

Figure~\ref{fig:SRMphi} depicts the parametric gain on ETMX for modes 37, 41 and 257 as a function of position of the signal-recycling mirror (SRM), expressed as tuning relative to RSE. Each mechanical mode is modelled at its \Comsol determined frequency (see figure~\ref{fig:maps}). Detuning the SRC causes a minor alteration to the operating point of the interferometer (see Section~\ref{sec:Method}); however, actively tuning the interferometer linear degrees of freedom to maintain operating point did not significantly change our results. 

For mode 37, we find a broad peak in parametric gain, resulting in instability for the nominal tuning and an increase in parametric gain for negative detunings. The SRM and Input Test Mass in the X-arm (ITMX) can be viewed as a compound mirror with a frequency-dependent reflectivity determined by the phase accumulated in the SRC. Changes to the position of the SRM therefore alter the effective reflectivity of ITMX as seen by the higher order optical modes. 

\begin{figure}[t]
\centering
    \begin{subfigure}[b]{0.39\textwidth}
        \includegraphics[width=\textwidth]{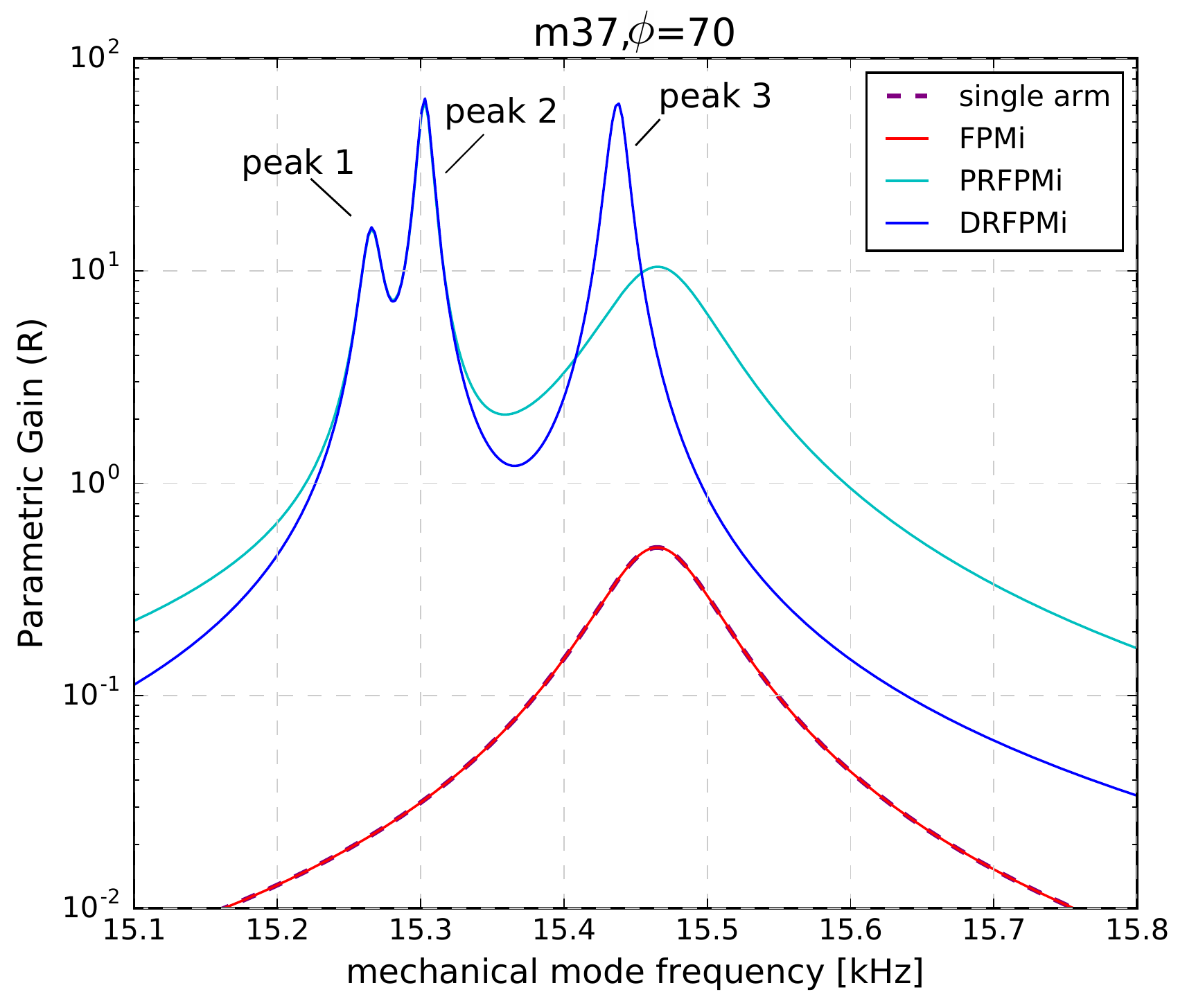}
        \caption{}
        \label{fig:phigif}
    \end{subfigure}
    ~ 
    \begin{subfigure}[b]{0.4\textwidth}
        \includegraphics[width=\textwidth]{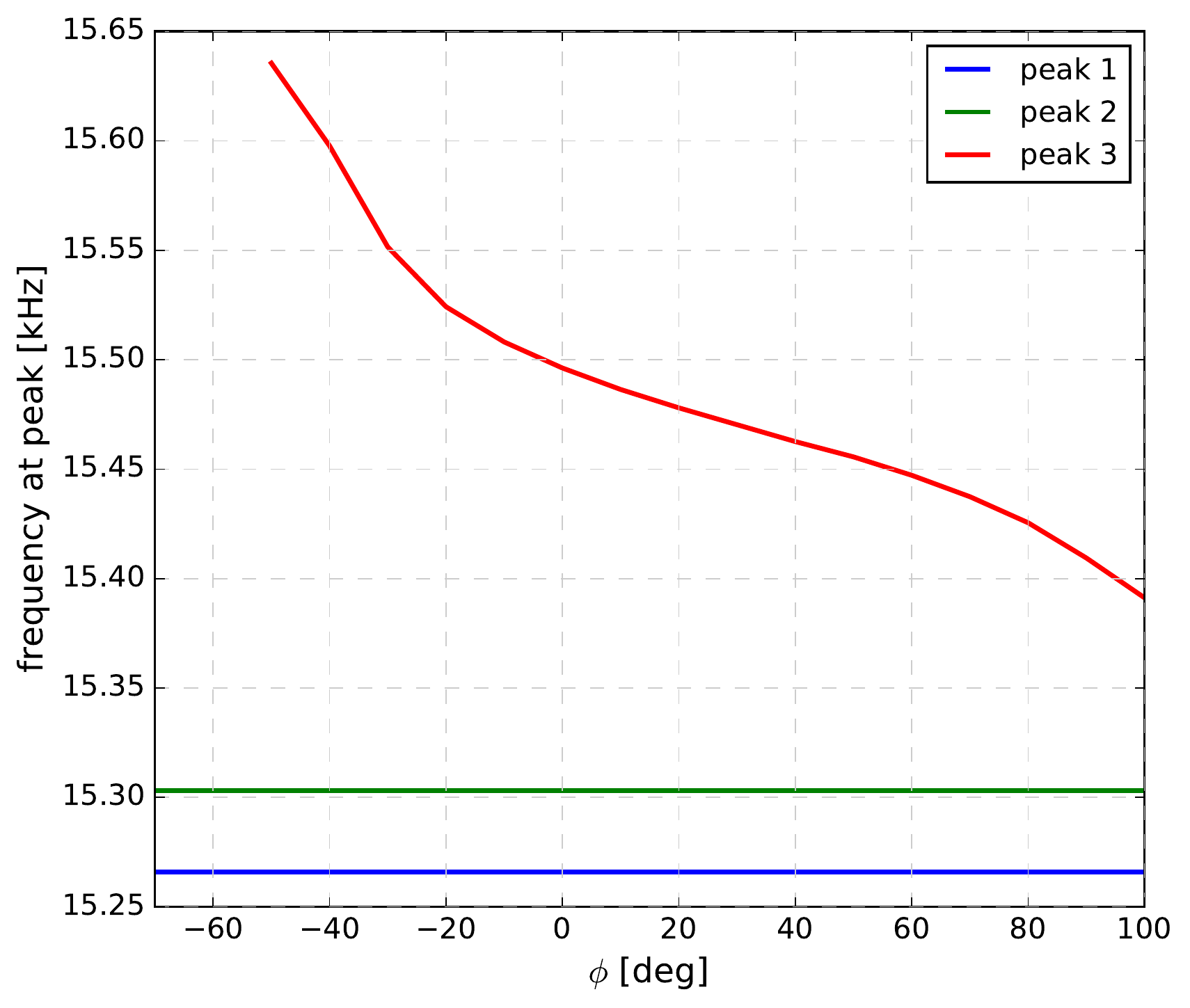}
        \caption{}
        \label{fig:phitracking}
    \end{subfigure}
\caption{Parametric gain of mode 37 as a function of mechanical mode frequency, plotted for different choices of SRM position, expressed as SRC tuning relative to RSE. In \ref{fig:phigif} we show the example $\phi=+70^\circ$. On comparison to figure~\ref{fig:builddown-as} we see that the differential peak has shifted in frequency and amplitude, while the common peaks introduced by the PRC are not affected. In \ref{fig:phitracking} we plot the mechanical mode frequencies resulting in peaks of parametric gain as the SRM position changes. This confirms that only the peak identified as `differential' is affected by SRC tuning. An animated gif of figure~\ref{fig:phigif} showing all SRM positions is available online.}
\label{fig:SRMphigif}
\end{figure}

Figure~\ref{fig:SRMphigif} directly tracks the interferometer response to the tuning of the SRC.  The animation in \ref{fig:phigif} plots parametric gain as a function of mechanical mode frequency for different choices of SRC tuning, including depicting the power-recycled and single cavity responses for reference. Figure~\ref{fig:phitracking} extracts the mechanical frequency corresponding to each of the three found peaks. We see that the common mode peaks are left unaltered, while the frequency of the differential peak doublet strongly depends on the position of the signal-recycling mirror. 

The differential mode peak frequency is not depicted for tunings of -70$^\circ$ and -60$^\circ$ since at these values the peak coincides with the common mode peaks and cannot be resolved. In this sense, the differential mode peak appears suppressed. We note that although mode 37 falls within the differential peak, and is influenced by the SRC, other mechanical modes will fall within the common mode peaks, and thus be independent of parameters in the SRC, as shown in figure~\ref{fig:phitracking}. 

Figure~\ref{fig:SRMphi} also shows the behaviour of the parametric gain as a function of SRC tuning for modes 41 and 257, both at full design power. For mode 41, we find that the mode is stable for all planned SRC tunings. However, we find that mode 257 could become unstable if the operational mode were to be switched away from RSE to positive detunings. The set of mechanical modes that could result in parametric instability can change depending on the SRC tuning. 

\subsection{Gouy Phase}
Proposed upgrade plans for Advanced LIGO include replacing the SRM with another mirror of a different curvature. This would alter the Gouy phase accumulated in the SRC and therefore the optical gain of higher order optical modes in this cavity. 
We find that changes in the Gouy phase have the same effect as SRC tuning for mode 37. 

\begin{figure}[b]
\centering
    \begin{subfigure}[b]{0.39\textwidth}
        \includegraphics[width=\textwidth]{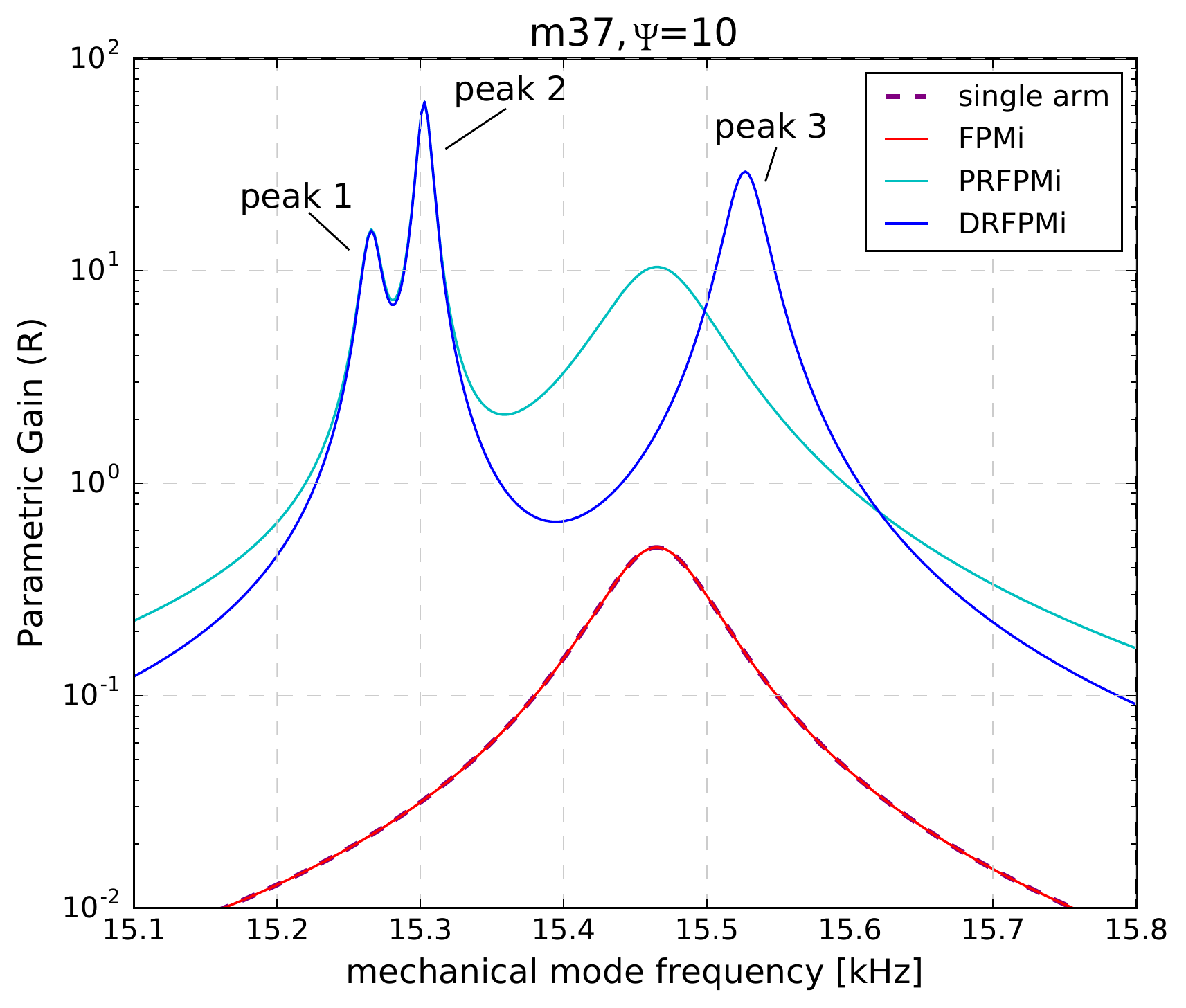}
        \caption{}
        \label{fig:gouygif}
    \end{subfigure}
    ~ 
    \begin{subfigure}[b]{0.4\textwidth}
        \includegraphics[width=\textwidth]{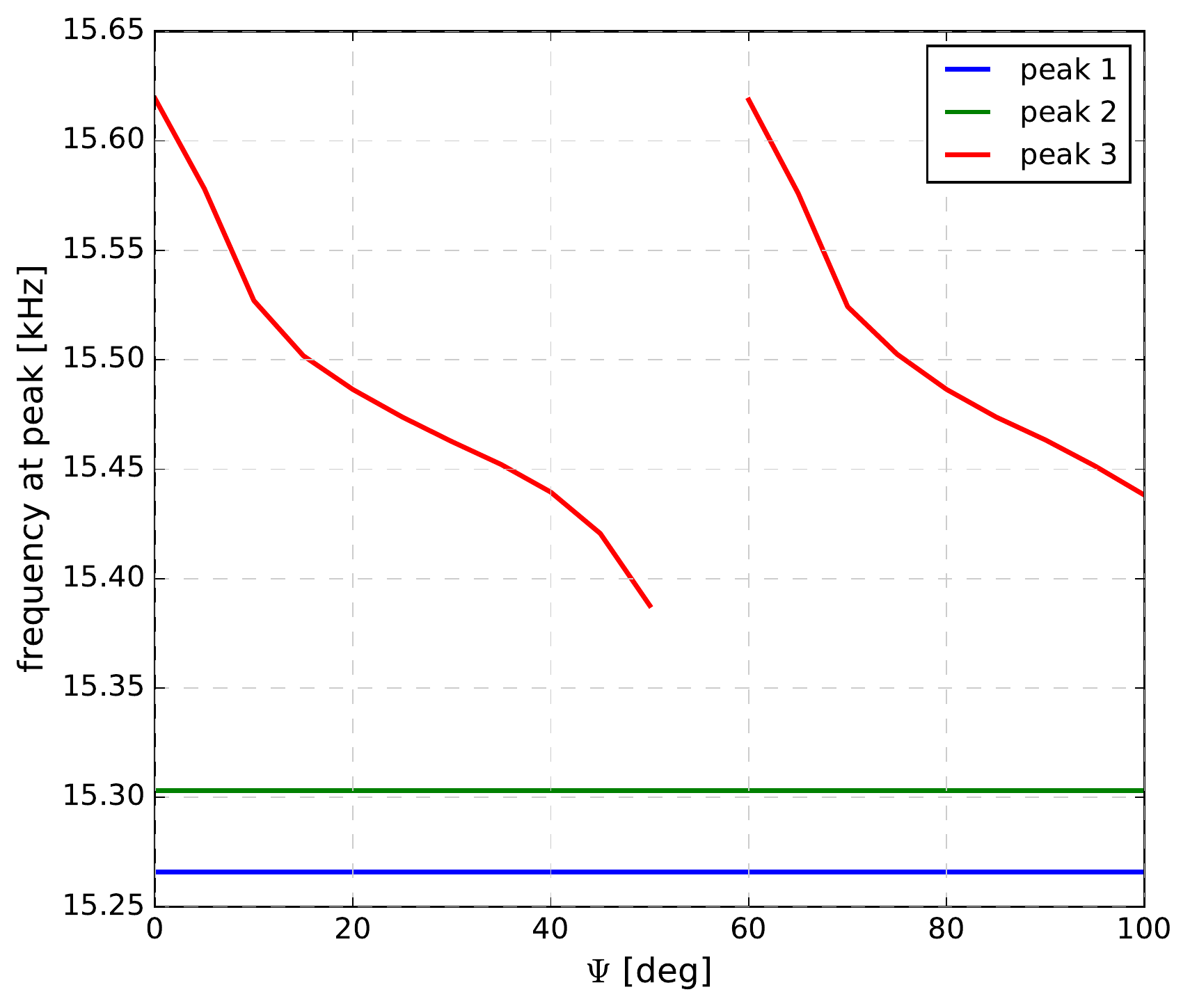}
        \caption{}
        \label{fig:gouytracking}
    \end{subfigure}
\caption{ \ref{fig:gouygif} shows the parametric gain of mode 37 as a function of mechanical mode frequency for the case of $\Psi= +10^\circ$ and is available as an animated gif online for different values of Gouy phase. \ref{fig:gouytracking} extracts the peak frequencies at each value of $\Psi$. As in figure~\ref{fig:SRMphigif}, we see that only the differential peak is affected by changes to the accumulated Gouy phase in the SRC. The missing points in the trace for $\Psi\sim50-60^\circ$ correspond to values for which the third peak cannot be resolved.}
\label{fig:SRMgouygif}
\end{figure}

In figure~\ref{fig:SRMgouygif}, we directly set the value of the Gouy phase accumulated in a single pass through the space between the SRM and telescope mirror SR2, $\Psi$ (see figure ~\ref{fig:OptLayout})\footnote{This allows us to mimic the effect of changing the radius of curvature of one of these mirrors without additional design time to re-mode-match the model.}. As in figure~\ref{fig:SRMphigif}, we see that only the differential mode doublet is affected by the change, and that the frequency range for which $\mathbb{R}>1$ changes with Gouy phase, to the point where we can suppress the differential peak of mode 37. Note that the periodicity of this behaviour is double that of the SRC tuning case; this is due to setting the \textit{one-way} rather than round-trip phase. 

\subsection{Consequences for Advanced LIGO Interferometers}
\begin{figure}[htb]
\centering
    \begin{subfigure}[b]{0.4\textwidth}
        \includegraphics[width=\textwidth]{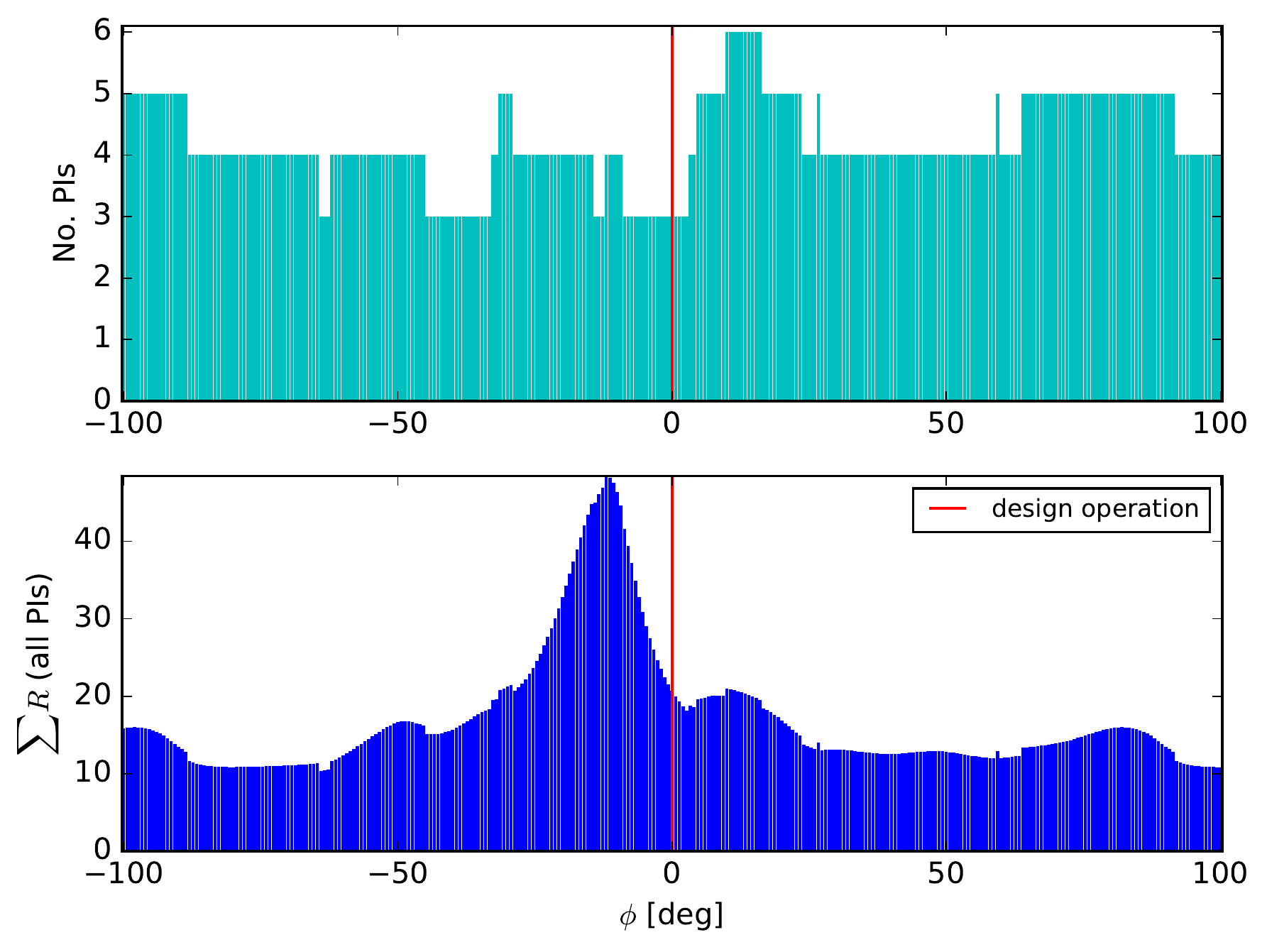}
        \caption{Tuning}
        \label{fig:TotPhi}
    \end{subfigure}
    ~ 
    \begin{subfigure}[b]{0.4\textwidth}
        \includegraphics[width=\textwidth]{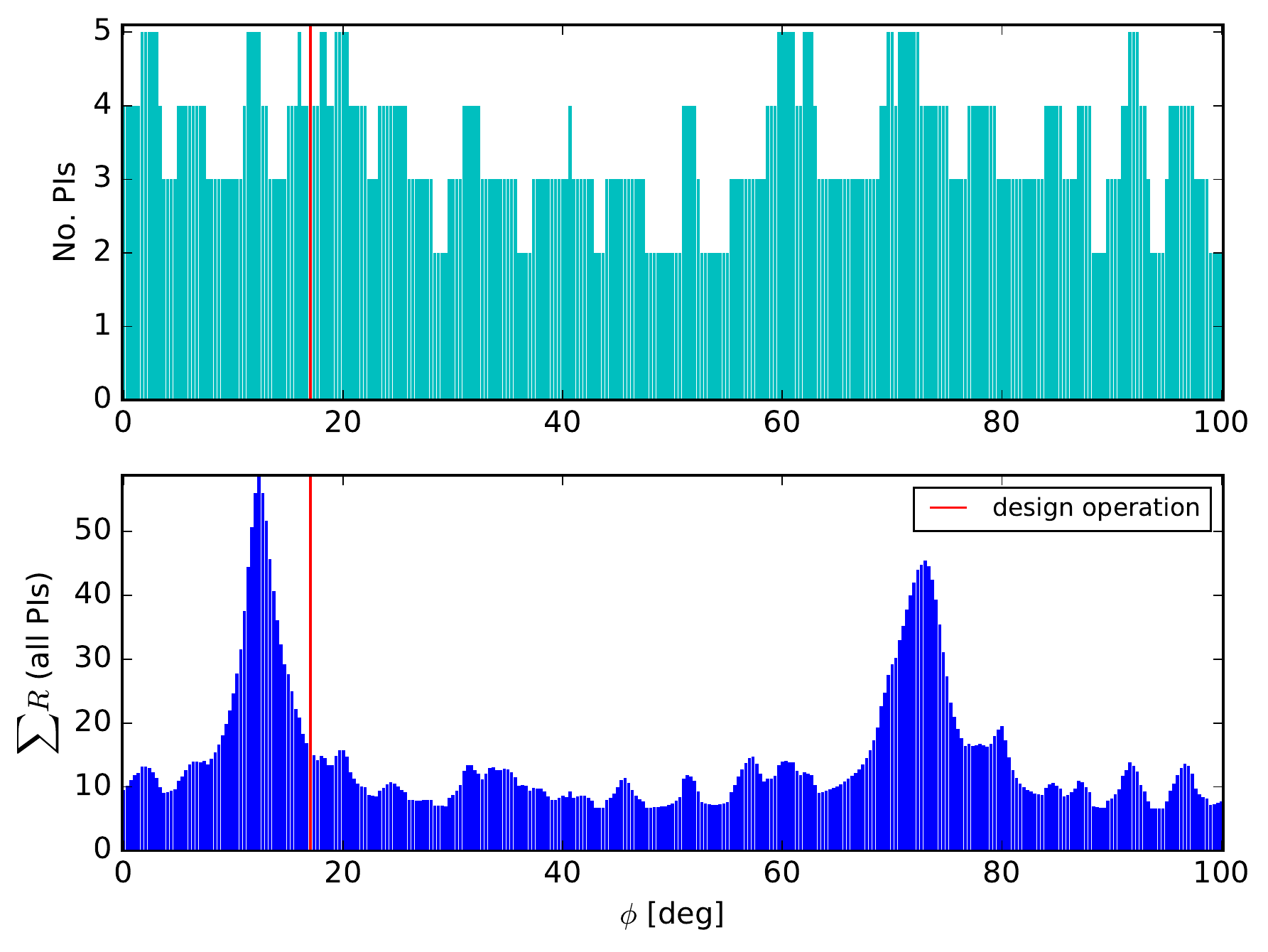}
        \caption{Gouy phase}
        \label{fig:TotGouy}
    \end{subfigure}
\caption{Total number of unstable modes and summed parametric gain for different choices of detuning, $\phi$ (\ref{fig:TotPhi}) and Gouy phase, $\Psi$ (\ref{fig:TotGouy}). The design operating point of the interferometer is marked for reference.}
\label{fig:TotPIs}
\end{figure}

We have calculated the parametric gain of 800 mechanical eigenmodes for discrete tunings in $-100^\circ \geq \phi \geq +100^\circ$ and Gouy phases in $0^\circ \geq \Psi \geq +100^\circ$. This is summarised in figure~\ref{fig:TotPIs}. In all cases the mechanical modes are modelled at their calculated \Comsol frequency and the power circulating in the X-arm cavity is $750$kW. In the case of SRC tuning, we find a minimum of 1 and maximum of 6 unstable modes in our model. The summed $\mathbb{R}$ plots are dominated by the gain of just one or two modes, as can be seen by comparing the shape of the lower plot in figure~\ref{fig:TotPhi} to the trace for mode 37 in figure~\ref{fig:SRMphi}. We also find that the current tuning of the SRC sits at a local minimum in terms of number of modes, however the gain of this mode is relatively high when compared to other minima in the upper trace: at $\phi  \simeq -40^\circ$ we are closer to suppressing all modes. As expected from comparing figures~\ref{fig:SRMphigif} and \ref{fig:SRMgouygif}, this behaviour is also shown for the case of Gouy phase changes. 

Relative to the design operating point as show in figure~\ref{fig:TotPIs}, the number of unstable modes increases for small positive SRC detuning but the gain of a single PI increases for the equivalent negative detuning. The position and radii of curvature of the mirrors in the signal-recycling cavity alter the parametric gain of mechanical modes and have a strong influence on which PIs will appear. This may have consequences for the current mitigation scheme, such as targeting new mode shapes or requiring stronger actuation to damp higher gain modes. However as seen in our results, the number of additional PIs that could arise in the current configuration are limited. 


\section{Conclusion}

We have used \Finesse to study parametric instabilities in the context of the full Advanced LIGO design. The dual-recycled configuration of the interferometer greatly expands the parameter space which determines the resulting optical gain of the system. In particular, we have shown that parameters outside the Fabry-Perot arm cavities can also affect the parametric gain of a mechanical mode, to the extent that the number and gain of unstable modes may change.

By contrasting figures~\ref{fig:RvRoCs} and \ref{fig:SRMphi}, the complexity of the picture is clear: while the instability of mode 41 is minimally affected by SRC tuning and strongly affected by changes in RoC, the reverse appears true of mode 257, and mode 37 is affected by both. The list of important optical parameters is therefore extensive, and all will influence the likely number of PIs that will affect gravitational wave detectors as the operating power increases. 

For the parameters in the Advanced LIGO design, we find that the tuning and Gouy phase accumulated in the signal-recycling cavity will influence the total number of parametric instabilities, and the gain of these modes. For differential modes parametric instability depends on properties in the SRC, while for common modes instability depends on the PRC. Therefore if parameters in the SRC are to be changed, a PI mitigation scheme based on per-mode damping is expected to remain effective for common mode PIs, but may require changes for differential modes. 

\section{Acknowledgements}
This work was supported by the Science and Technology Facilities Council Consolidated Grant (number ST/N000633/1) and H.M. is supported by UK Science and Technology Facilities Council Ernest Rutherford Fellowship (Grant number ST/M005844/11).

\appendix
\section{Parametric Instability}\label{app:PIintro}

\begin{figure}[h]
\begin{center}
\includegraphics[width=0.4\textwidth]{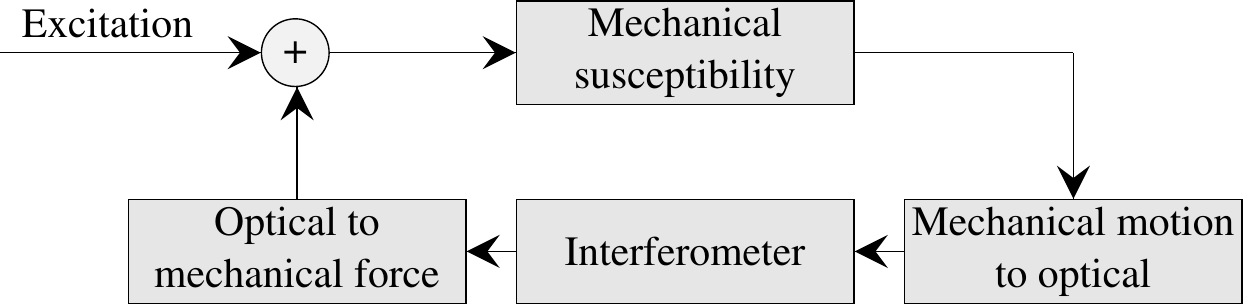}
\caption{Schematic depiction of parametric instability as a feedback process. A mechanical excitation of the optic scatters light from the fundamental optical mode into higher order optical modes. These circulate and interact with the fundamental optical  field and mechanical modes via radiation pressure. If the strength of the interaction and the optical gain of the resulting modes are sufficiently high, the mechanical mode may be unstable: a parametric instability.}
\label{fig:PIFB}
\end{center}
\end{figure}

Throughout this work we consider the linear interaction between the optical field and a vibrational mode of a suspended mirror within the interferometer. As explained by \cite{Evans2010665} these parametric instabilities can be described as a feedback system, depicted schematically in figure~\ref{fig:PIFB}. An excitation acts on a particular vibrational mode of the mirror, causing the reflecting surface of the optic to move. The incoming optical field (`pump') is phase modulated on reflection, resulting in scattering of the optical field into higher order optical modes (HOMs) at sideband frequencies determined by the frequency of the mirror surface motion. We describe the resulting optical modes in the Hermite-Gauss (HG) basis. The lower sideband corresponds to the Stokes mode - energy coupling from the optical field into the mechanical oscillation -  while the upper sideband corresponds to the reverse process, the anti-Stokes mode. The strength of the interaction depends on the spatial overlap of the mechanical mode, $m$, with the incoming and scattered optical field modes, denoted $B_{m,n}$ for the $n$th optical mode \cite{Evans2010665}. The resulting optical fields then propagate through the interferometer, where they may be amplified or suppressed depending on their frequency and the particular configuration of the interferometer. On returning to the mirror, radiation pressure will act on the mirror surface. If the upper sideband dominates, the motion will be damped; if instead the lower sideband dominates, energy is coupled out of the optical field into the mirror and it `rings up', resulting in a parametric instability.

The figure of merit for determining the stability of a vibrational mode, $m$, in an interferometer is called the \textit{parametric gain}, $\mathbb{R}$, where $\mathbb{R} \geq 1$ corresponds to an instability. In the case of a single dominant incident field of wavelength $\lambda$ and power $P$, this is given by:

\begin{equation}
\mathbb{R}_m = \frac{8 \pi Q_m P}{M \omega_m^2 c \lambda} \sum_{n=0}^{\infty}\Re[G_n]B^2_{m,n}.
\label{eq:R}
\end{equation}
where $0\geq B_{n,m} \geq 1$ is as above, $c$ is the speed of light, $M$ is the mass of the mirror, and vibrational mode $m$ has angular resonant frequency $\omega_m$ and quality factor $Q_m$. For the fused silica LIGO test masses $M=$~40\,kg, $\omega_m \sim 10\times 2\pi$kHz and $Q_m\sim10^7$. The incident optical field has $\lambda =$ 1064nm and $P=750$kW.  $G_n$ is the optical transfer function of the $n$th HOM through the interferometer and back to the mirror. Changes to the interferometer configuration will therefore affect which vibrational modes are likely to become unstable in an interferometer. The linear dependence of $\mathbb{R}$ on incident power indicates that vibrational modes that are stable for low powers may become unstable once the detectors are upgraded to full design power and sensitivity.

\section{A `Forest of Modes'}
\label{app:ForestDRMi}

\begin{figure}[htb]
\centering
    \begin{subfigure}[b]{0.5\textwidth}
        \includegraphics[width=\textwidth]{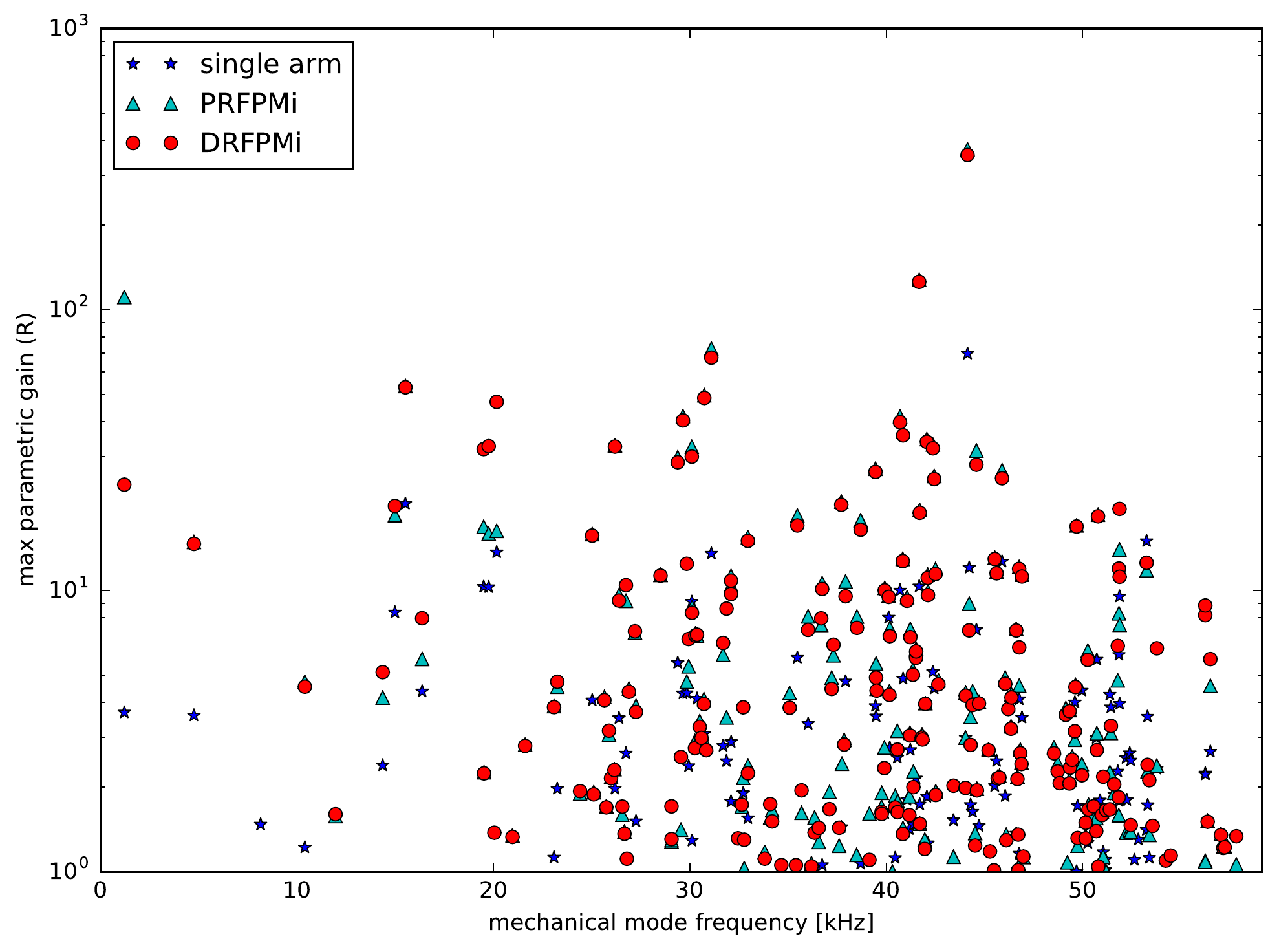}
        \caption{}
        \label{fig:forest-BD}
    \end{subfigure}
    ~ 
    \begin{subfigure}[b]{0.3\textwidth}
        \includegraphics[width=\textwidth]{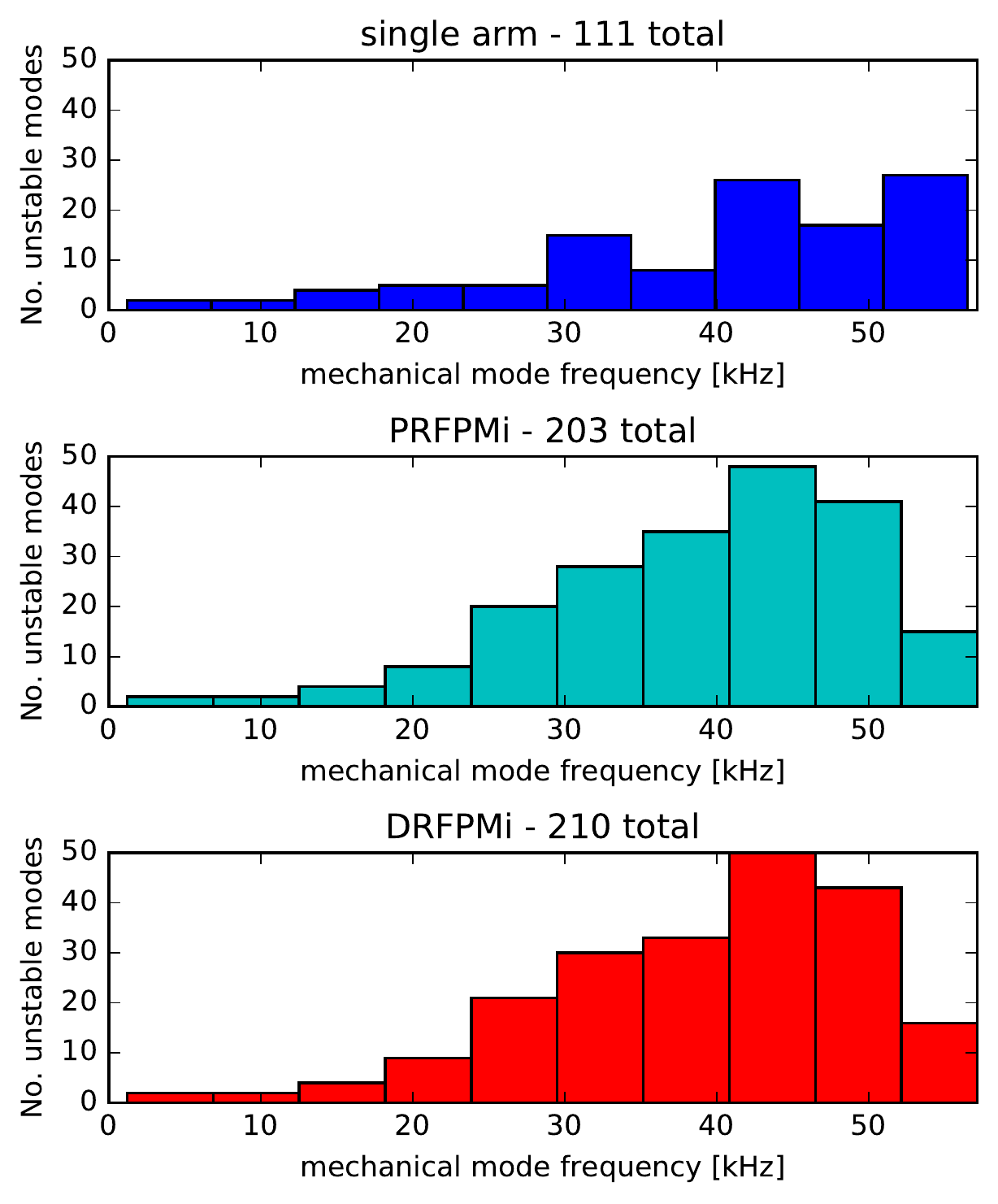}
        \caption{}
        \label{fig:forest-BD-hist}
    \end{subfigure}
\caption{A forest of PIs: each data point relates to a specific mechanical mode that has positive parametric gain within $\pm$2 kHz of its \Comsol computed resonant frequency. 3 cases are depicted: a single arm cavity, PRFPMi, and DRFPMi interferometers. We find that the DRFPMi configuration could produce twice the number of unstable modes when compared to the single cavity.}
\label{fig:BDForest}
\end{figure}

Figure~\ref{fig:BDForest} illustrates the importance of including the full DRFPMi interferometer in PI studies for LIGO. Building the interferometer in stages as described in section~\ref{sec:Builddowns}, we plot all mechanical modes up to 60kHz that are found to be unstable within 2~kHz of their \Comsol frequency. Each point then marks the peak value of $\mathbb{R}$ found for each eigenmode. This allows for inaccuracies in our simple mechanical model and a range of interferometer parameters,  creating a 'worst case scenario' for PIs at LIGO. Critically, we find that the dual-recycled interferometer could suffer from twice the number of PIs when compared to the single cavity case. 
Note also that this plot refers to PIs exclusively due to mechanical modes in ETMX. Modes from different test masses, and any cross-coupling between these, are not considered in this study. Therefore the total number of PIs could at worst be quadruple that depicted.

\begin{figure}[htb]
\begin{center}
\includegraphics[width=0.4\textwidth]{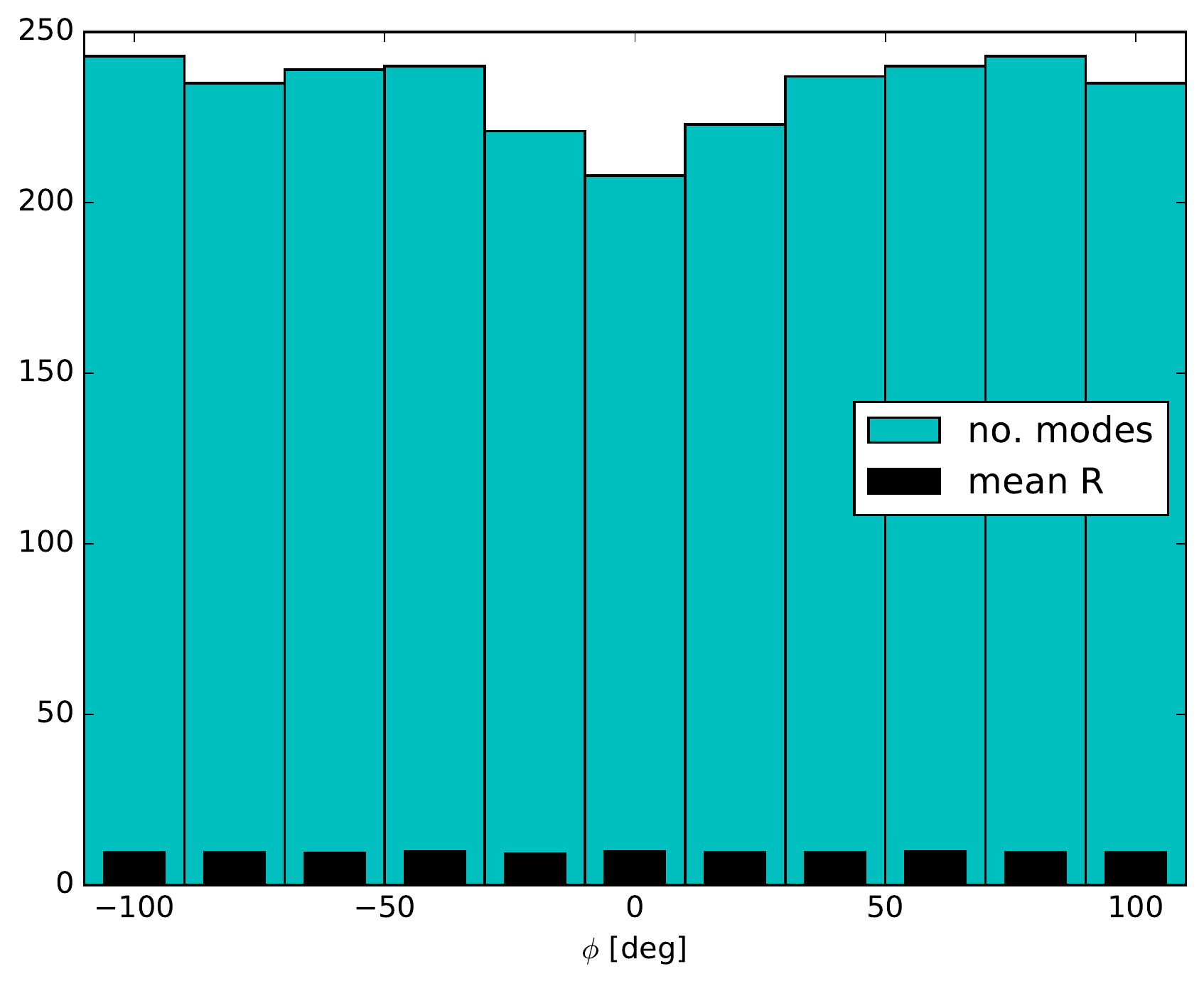}
\caption{Total number of modes that are unstable within $\pm$2kHz of their \Comsol frequency. The influence of the SRC is diluted here since this range of mechanical frequencies also allows most modes to resonate via a common rather than differential mode.}
\label{fig:SRMbars}
\end{center}
\end{figure}

This approach has also been used to study the influence of SRC tuning, as shown in figure~\ref{fig:SRMbars}. We find that in a `worst case scenario', whereby all interferometer parameters combine to maximise the number of PIs, the influence of SRC tuning on the total number of PIs is diluted. By allowing the mechanical frequency to sweep over a 4kHz range, peaks in parametric gain due to both power- and signal-recycling are included, and a value of $\mathbb{R}>1$ anywhere in this range is treated as a count of 1 unstable mode. Since the majority of modes are able to resonate in the PRC (given an appropriate choice of mechanical frequency), changing the tuning of the SRC just influences the minority of modes that are only resonant via the SRC.

\newpage
\bibliographystyle{myunsrt}
\bibliography{PI-LIGO}

\end{document}